\documentclass[prb,twocolumn,showpacs,preprintnumbers,superscriptaddress,amsmath,amssymb]{revtex4}  
\usepackage{graphicx}
\usepackage{dcolumn}

\begin{document}

\title{A neutron scattering study of the under-doped Ba$_{1-x}$K$_{x}$Fe$_{2}$As$_{2}$, x=0.09 and 0.17 self-flux grown single crystals and the universality of the tricritical point}

\author{C. R. Rotundu}
\email[Corresponding author: ]{CRRotundu@lbl.gov}
\affiliation{Materials Sciences Division, Lawrence Berkeley National Laboratory, Berkeley, CA 94720, USA}
\author{W. Tian}
\affiliation{Ames Laboratory and Department of Physics and Astronomy, Iowa State University, Ames, IA 50011, USA}
\author{K. C. Rule}
\affiliation{Helmholtz-Zentrum Berlin, Berlin, Germany}
\author{T. R. Forrest}
\affiliation{Department of Physics, University of California, Berkeley, CA 94720, USA}
\author{J. Zhao}
\affiliation{Department of Physics, University of California, Berkeley, CA 94720, USA}
\author{J. L. Zarestky}
\affiliation{Ames Laboratory and Department of Physics and Astronomy, Iowa State University, Ames, IA 50011, USA}
\author{R. J. Birgeneau}
\affiliation{Department of Physics, University of California, Berkeley, CA 94720, USA}
\affiliation{Department of Materials Science and Engineering, University of California, Berkeley, CA 94720, USA}
\date{\today}

\begin{abstract}

We present a combination of elastic neutron scattering measurements in zero and 14.5 T and magnetization measurements in zero and 14 T on under-doped superconducting Ba$_{1-x}$K$_{x}$Fe$_{2}$As$_{2}$ x=0.17, and the same measurements in zero field on a non-superconducting crystal with x=0.09. The data suggest that the under-doped materials may not be electronic phase separated but rather have slightly inhomogeneous potassium doping. The temperature dependence of the magnetic order parameter (OP) below the transition of the sample with x=0.09 is more gradual than that for the case of the un-doped BaFe$_{2}$As$_{2}$, suggesting that this doping may be in the vicinity of a tricritical point. We advance therefore the hypothesis that the tricritical point is a common feature of all superconducting 122s. For the x=0.17 sample, while T$_{c}$ is suppressed from $\approx$17 K to $\approx$8 K by a magnetic field of 14 T, the intensity of the magnetic Bragg peaks (1 0 3) at 1.2K is enhanced by 10$\%$ showing competition of superconductivity (SC) and antiferromagnetism (AFM). The intensity of the magnetic Bragg peaks (1 0 3) in the (T$_{c}$, T$_{N}$) temperature interval remain practically unchanged in 14.5 T within a 10$\%$ statistical error. The present results are discussed in the context of the existing literature.
\end{abstract}


\maketitle

\section{Introduction}

High temperature superconductivity (HTSc) in the iron pnictides, with a T$_{c}$ as high as 55 K for the case of SmFeAsO$_{1-\delta}$ and SmFeAsF$_{x}$O$_{1-x}$ \cite{Ren,Ren2}, is  one of the most perplexing discoveries of the decade in the field of condensed matter physics. The 122 series (AFe$_{2}$As$_{2}$, A = Ba, Sr, Ca, Eu) is of great interest since it is an oxygen-free HTSc. Superconductivity in the 122s can be induced by doping in any of the three atomic sites \cite{Rotter2,Shirage,Sasmal,Sefat,Leithe,Li,Ren3,Schnelle,XLWang,NNi}. The hole-doping achievable through chemical substitution with either K \cite{Rotter2}, Na \cite{Shirage} or Cs \cite{Sasmal} in the atomic site $A$ can give a T$_{c}$ as high as 39 K in the case of Ba$_{0.55}$K$_{0.45}$Fe$_{2}$As$_{2}$ \cite{Rotter2}.
The antiferromagnetic (spin-density wave) and structural (tetragonal to orthorhombic) transitions that are near-coincident in the parent compounds \cite{Rotter1,Huang} are concomitantly and gradually suppressed upon doping. Although in the electron-doped BaFe$_{2}$As$_{2}$ the two transitions separate with doping \cite{Canfield}, it seems that there are examples pointing otherwise, as in the case of the isovalent ruthenium doped BaFe$_{2}$As$_{2}$ \cite{Thaler,Kim2} and the case of electron-doped SrFe$_{2(1-x)}$Co$_{2x}$As$_{2}$ \cite{Gillett}. The last is surprising if we consider the result on Sn-flux grown Ca(Fe$_{1-x}$Co$_{x}$)$_{2}$As$_{2}$ crystals \cite{Harnagea} for which the two transitions are clearly separated. In the case of potassium (hole) doped BaFe$_{2}$As$_{2}$ \cite{Avci} the question of concomitant or separated transitions remains controversial. While powder neutron diffraction data argue for concomitant magnetic and structural transitions across the whole series \cite{Chen,Avci}, heat capacity on Sn-flux grown Ba$_{0.84}$K$_{0.16}$Fe$_{2}$As$_{2}$ single crystals shows two distinctive peaks attributed by the authors to the magnetic and structural phase transitions, respectively\cite{Urbano}. In most cases the source of contradictory results appears to be connected to issues of sample quality. It has been pointed out that the proper flux to grow the 122s is FeAs \cite{Wang,Greg,Su}, as other fluxes contaminate the sample with flux element inclusions, with a consequent impact on the physical properties. Neutron diffraction on powder BaFe$_{2}$As$_{2}$ \cite{Huang} determined a first-order structural and magnetic transition. Complementary high resolution X-ray diffraction and heat capacity measurements on high quality BaFe$_{2}$As$_{2}$ crystals revealed a 1$^{st}$ order magnetic transition preceded by a structural transition that starts as a 2$^{nd}$ order transition at a slightly higher temperature but with a first order jump in the orthorhombic distortion coincident with the first order magnetic transition\cite{Rotundu1,Kim}. For the electron-doped BaFe$_{2(1-x)}$Co$_{2x}$As$_{2}$ it has been shown recently that the magnetic transition order changes upon doping from 1$^{st}$ to 2$^{nd}$ through a tricritical point \cite{Kim,Rotundu3}, which is believed to be relevant to the superconductivity phenomenon itself \cite{Giovannetti}. For this series the structural transition is 2$^{nd}$ order. This seems to be different for the case of polycrystalline hole-doped Ba$_{1-x}$K$_{x}$Fe$_{2}$As$_{2}$ of C. Avci $\emph{et al.}$ \cite{Avci} for which both the magnetic and structural transitions are 1$^{st}$ order over the entire doping range. An early report on Sn-flux grown K-doped BaFe$_{2}$As$_{2}$ revealed an electronic phase separated material \cite{Park}. A more recent atom probe tomography study on self-flux grown underdoped Ba$_{0.72}$K$_{0.28}$Fe$_{2}$As$_{2}$ provides evidence for a mixed scenario of phase coexistence and phase separation originating from variation of the dopant atom distributions \cite{Yeoh}.
In the present article we report complementary zero and 14.5 T elastic neutron scattering and zero and 14 T magnetization measurements on underdoped non-SC x=0.09 and SC x=0.17 (T$_{c}$$\approx$17 K). For the non-SC x=0.09 sample the AFM transition is sharp (width within 1 K), consistent with a weakly 1$^{st}$ order transition, and the temparature dependence of the magnetic order parameter (OP) squared is more gradual than that for the case of the parent BaFe$_{2}$As$_{2}$. This possibly indicates proximity to a tricritical point.
For the higher doping sample x=0.17, the transition presents a distribution of T$_{N}$s due to a slight variation of the potassium dopant, leading to a rounding of the transition of about 6 K. This rounding makes it difficult to differentiate between first and second order behavior of the transition. Our neutron data show that, although the SC under-doped x=0.17 sample has a SC volume fraction of $\approx$40$\%$, the downturn in the AFM order parameter below T$_{c}$ and its enhancement in magnetic field provide possible evidence for microscopic coexistence of AFM and SC, similar to the case of the electron-doped 122s\cite{Fernandes}.

\section{Experimental Procedure}

Single crystals of Ba$_{1-x}$K$_{x}$Fe$_{2}$As$_{2}$ with potassium dopings $x$ of 0.09, 0.17, 0.41, and 0.45 \cite{K41} were synthesized by a self-flux method with details given in an earlier report \cite{Rotundu2}. Samples with x=0.41 and 0.45 show no sign of AFM ordering, and are fully superconducting with T$_{c}$=38 K \cite{Rotundu2} and 39 K, $\Delta$T$_{c}$=2 K, respectively. The potassium doping has been extrapolated by comparing the $c$ lattice determined from room temperature neutron data with $c$ vs. $x$ (K doping) data on crystals of H. Lou $\emph{et al.}$ \cite{Luo} (Fig. 1). This data agree with $c$ vs. $x$ of the polycrystalline samples \cite{Johrendt}. Crystals of SrFe$_{2-x}$Ni$_{x}$As$_{2}$, x=0.155 were grown by self-flux method as well \cite{Paglione2} and the precise Ni doping value was determined by inductively coupled plasma analysis.

\begin{figure}[h]
\begin{center}\leavevmode
\includegraphics[width=1.1\linewidth]{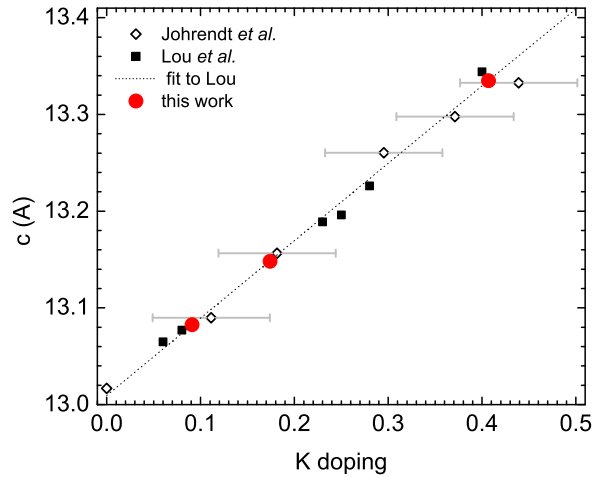}
\caption {$c$ lattice versus $x$ potassium doping, literature and present study data.}
\label{fig2}\end{center}\end{figure}

Magnetization measurements on the K-doped samples were carried out using a Magnetic Property Measurement System (MPMS) and a Physical Property Measurement System (PPMS) from Quantum Design$\textsuperscript{\textregistered}$. Resistivity measurements on the Ni-doped SrFe$_{2}$As$_{2}$ were performed in the PPMS. Zero field neutron diffraction measurements were performed at Oak Ridge National Laboratory (ORNL) with the High Flux Isotope Reactor's HB-1A triple axis spectrometer, using a horizontal collimation 40' -- 40' - $sample$ -- 40' -- 68' and fixed energy E$_{i}$=14.6 meV. The samples studied had K concentrations of x=0.09, 0.17, and 0.41 with masses of 19.5 mg, 45 mg, and 71.5 mg, respectively. In order to assess further sample quality, rocking cuvres of the (008) Bragg peak were recorded. For the x=0.09 sample the rocking curve showed two peaks that were separated by approximately 0.7$^{\circ}$. Fitting these peaks to a Lorentizan squared profile gave FWHMs of 0.60$^{\circ}$ and 0.70$^{\circ}$. For the x=0.17 sample, the rocking curve of the same peak gave one main peak with a FWHM of 0.77$^{\circ}$.

Neutron diffraction measurements of the x=0.17 sample of 50 mg in zero field and 14.5 T were performed at the Helmholtz-Zentrum Berlin (HZB) with a configuration of 60' -- 20' - $sample$ -- 20' and fixed E$_{f}$=5.0 meV. The rocking curve of the (002) Bragg peak showed two peaks that were separated by 0.86$^{\circ}$, both peaks had a FWHM of 0.47$^{\circ}$.

For both ORNL and HZB neutron scattering experiments, the samples were mounted in a closed-cycle refrigerator and studied in the vicinity of the magnetic Bragg position Q$_{AFM}$=(1 0 3). For all magnetization and neutron scattering measurements, the magnetic field was parallel with the (a b) crystallographic plane.

\section{Results and Discussion}

\begin{figure}[h]
\begin{center}\leavevmode
\includegraphics[width=1.05\linewidth]{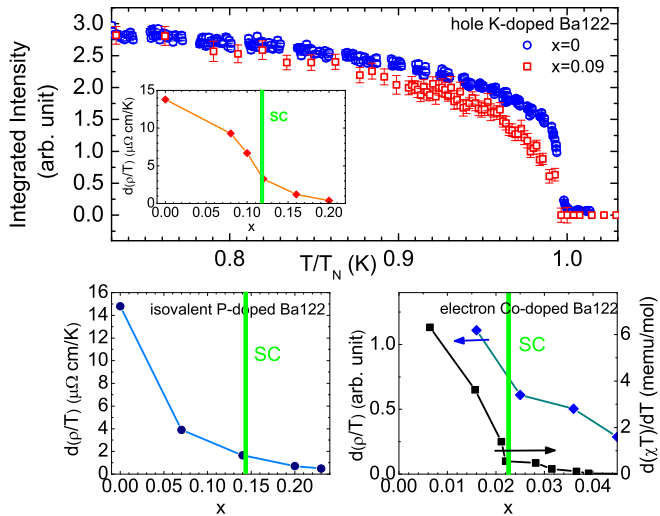}
\caption {The upper panel shows the integrated intensity of the (1 0 3) magnetic Bragg peak versus reduced temperature of the non-SC K-doped sample x=0.09 ($\Box$) plotted against the same data for the un-doped x=0 ($\circ$) (from Ref. 38). The inset shows the magnitude of the d$\rho$/dT peaks vs. $x$ potassium doping of the series extracted from Ref. 39. Similar d$\rho$/dT vs. $x$ data is plotted for the isovalent phosporus-doped BaFe$_{2}$As$_{2}$ in the lower left panel, and for the electron cobalt-doped in the lower right panel. The thick vertical lines indicate the doping corresponding to emergence of sueprconductivity in the series.}\label{fig4}\end{center}\end{figure}

The thermal evolution of the integrated intensity of the (1 0 3) magnetic Bragg peak in the non-SC x=0.09 sample is shown in Fig. 2, upper panel. In addition, the scattering from the same magnetic Bragg peak of the un-doped parent compound BaFe$_{2}$As$_{2}$ is also shown (data is taken from Ref. 38). The peak intensity scales like the magnetic OP squared. For the sample x=0.09 the N\'{e}el temperature is 136 K. The figure shows that the magnetic OP squared in the x=0.09 sample evolves in a much more gradual manner than Wilson $\emph{et al.}$'s data on x=0 \cite{Wilson}. Even so, there is a clear and sharp jump (within 1 K) of magnetic OP squared directly below the N\'{e}el temperature. This clearly shows that magnetic phase transition in this sample is still first order, albeit one that is weaker than the parent compound's. In critical phenomena language this corresponds to a slight increase in the effective critical exponent describing the temperature dependence of the OP below the first order transition. We speculate that x=0.09 may be close to a tricritical point, similar to the one found in Co-doped BaFe$_{2}$As$_{2}$ \cite{Kim,Rotundu3}. It has been shown for the electron Co-doped BaFe$_{2}$As$_{2}$ \cite{Rotundu3} that, around the tricritical point, the heat capacity $C$ and d($\chi$T)/dT vs. doping present a change from a more abrupt variation (characteristic of a 1$^{st}$ order transition), to a monotonic and much slower variation (characteristic of a 2$^{nd}$ order transition). For the electron Co-doped BaFe$_{2}$As$_{2}$ in the lower right panel of Fig. 2, the magnitude of the peaks d($\chi$T)/dT vs. doping reproduced from Ref. 28 is drawn in comparison with the magnitude of d$\rho$/dT \cite{Fisher} peaks as extracted from Ref. 40.
In the inset of the upper panel of Fig. 2 we plot the magnitude of the d$\rho$/dT peaks vs. $x$ of the same series, extracted from Ref. 39. The existence of an inflection point in the d$\rho$/dT vs. $x$ data may indicate a tricritical point at around x $\approx$ 0.12 for the hole K-doped system. As found for the case of the Co(electron)-doped BaFe$_{2}$As$_{2}$ \cite{Rotundu3}, this tricritical point in K(hole)-doped BaFe$_{2}$As$_{2}$ is in the near proximity of emergence of superconductivity 0.125 $\leq$ x $\leq$ 0.133 \cite{Avci}. Finally, the lower left panel shows magnitude of d$\rho$/dT peaks vs. phosphorus doping as extracted from Ref. 41 \cite{Kasahara}.

\begin{figure}[h]
\begin{center}\leavevmode
\includegraphics[width=1.1\linewidth]{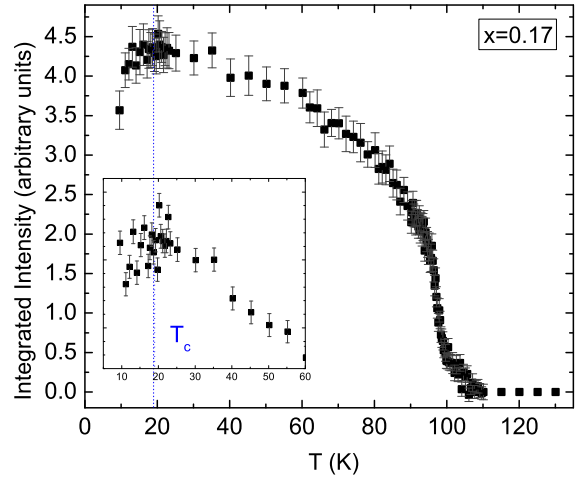}
\caption {Integrated intensity of the magnetic OP versus temperature for the sample x=0.17. The downturn of the OP below T$_{c}$ shows coexistence of AFM and SC. In the inset it is the sum of counts versus temperature near T$_{c}$. The vertical interrupted line is a guide to the eye for T$_{c}$. The rounded transition at T$_{N}$ is due to the slight K-doping inhomogeneity.}\label{fig4}\end{center}\end{figure}

Figure 3 shows the integrated intensity of the (1 0 3) magnetic Bragg peak versus temperature of the sample x=0.17. The downturn of the intensity below T$_{c}$ provides evidence for the  microscopic coexistence of AFM and SC. In the inset is shown the sum of counts versus temperature near T$_{c}$. The downturn of the magnetic OP is less pronounced than for the case of homologous superconducting electron doped BaFe$_{2(1-x)}$Co$_{2x}$As$_{2}$ \cite{Fernandes} because of the low superconducting volume fraction. The ``rounded'' N\'{e}el transition (over a $\approx$ 6 K temperature range) is due in part to a slight distribution in the potassium doping, and therefore this will give an averaged $\langle$T$_{N}$$\rangle$. If we assume that the effect is due solely to a spread in doping, then the $\approx$ 6 K wide transition corresponds on the phase diagram \cite{Johrendt} to a variation on potassium doping $x$ of about 2.5$\%$.
The presence in the sample of small fractions of material with slightly smaller values of potassium doping will result in a non-zero magnetic order parameter above $\langle$T$_{N}$$\rangle$ and SC critical temperatures below $\langle$T$_{c}$$\rangle$ (untraceable by means of resistivity and magnetization measurements).

\begin{figure}[h]
\begin{center}\leavevmode
\includegraphics[width=1.1\linewidth]{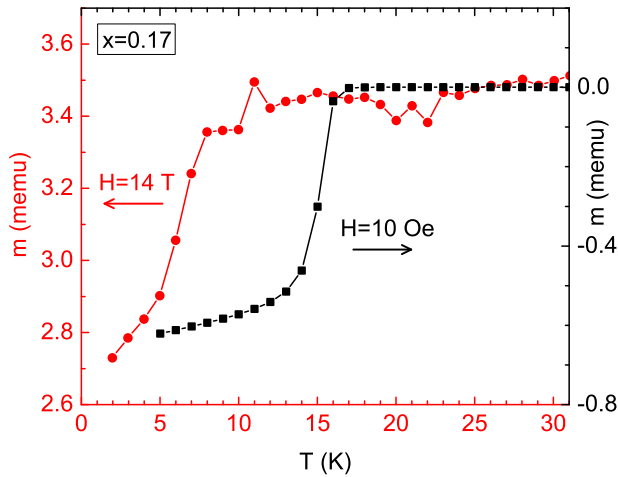}
\caption {Magnetization of the x=0.17 sample in H=10 Oe showing the onset of the diamagnetism at $\approx$17 K (right axis). In 14 T (left axis) T$_{c}$ decreases to $\approx$8 K.}\label{fig4}\end{center}\end{figure}

\begin{figure}[h]
\begin{center}\leavevmode
\includegraphics[width=1.05\linewidth]{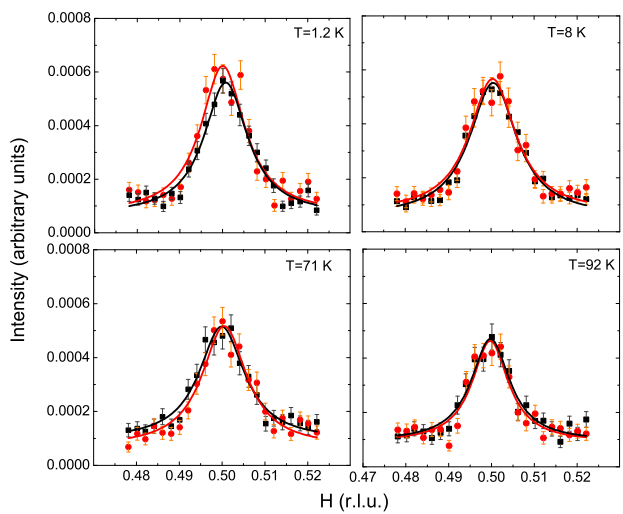}
\caption {Elastic neutron-scattering scans of the (1 0 3) magnetic Bragg peak at different temperatures: T=1.2, 8, 71, and 92 K of the sample x=0.17. For each temperature the zero field data are indicated with filled squares and data in 14.5 T with filled circles, and the curves are Lorentzian fittings.}\label{fig4}\end{center}\end{figure}

In order to investigate the effects of magnetic field on superconductivity, we have measured the magnetization of an x=0.17 sample. The onset of the diamagnetism as measured in 10 Oe is at $\approx$17 K (Fig. 4, right axis). The superconducting volume fraction is about 40$\%$. This value is considerably higher than the 23 $\%$ reported for a higher potassium doped Sn-flux grown sample \cite{Park}, and contrasts with the 98$\%$ value found by R. R. Urbano $\emph{et al.}$ \cite{Urbano} in their Sn-flux grown x=0.16. In a magnetic field of 14 T (Fig. 4, left axis) T$_{c}$ decreases to $\approx$8 K. This is expected since under-doped superconducting samples have a lower critical field H$_{c2}$ than those that are optimally doped, where the critical field was estimated to be above 75 T \cite{Altarawneh}.

Figure 5 shows elastic neutron-scattering scans of the (1 0 3) magnetic Bragg peak at different temperatures: T=1.2, 8, 71, and 92 K of the x=0.17 sample, in zero and 14.5 T. While for the 1.2 K the 14.5 T magnetic intensity is $\approx$10$\%$ higher than for the zero field, our data also show that a field of 14.5 T leaves the magnetic scattering practically unchanged within the errors for the (T$_{c}$, T$_{N}$) temperature range. This result certainly contrasts with the clear decrease of the magnetic intensity by $\approx$10$\%$ in 13.5 T reported on Sn-flux grown higher potassium doped BaFe$_{2}$As$_{2}$ (with T$_{c}$=32$\pm$1 K) of J. T. Park $\emph{et al.}$ \cite{Park}. Therefore our 14.5 T data on the hole K-doped BaFe$_{2}$As$_{2}$ is similar to the 10 T high resolution neutron data on the electron underdoped BaFe$_{1.92}$Ni$_{0.08}$As$_{2}$ (T$_{c}$=17 K) \cite{Wang2}. Here, bellow T$_{c}$ the intensity of the magnetic (1 1 3) peak is enhanced with $\approx$10$\%$, while above T$_{c}$ the intensity remains almost unchanged.
One experiment to test the interplay between AFM and SC would be to determine whether or not a high magnetic field induces AFM in an optimally doped sample (without any trace of static AFM in zero field). This is very difficult to apply to the case of the optimally K-doped BaFe$_{2}$As$_{2}$, as the critical field is over 75 T. Since for the case of the electron-doped 122s the critical field is much lower, we performed zero and in-field (14 T) resistivity measurements of optimally doped Ni-doped SrFe$_{2}$As$_{2}$.

\begin{figure}[h]
\begin{center}\leavevmode
\includegraphics[width=1.1\linewidth]{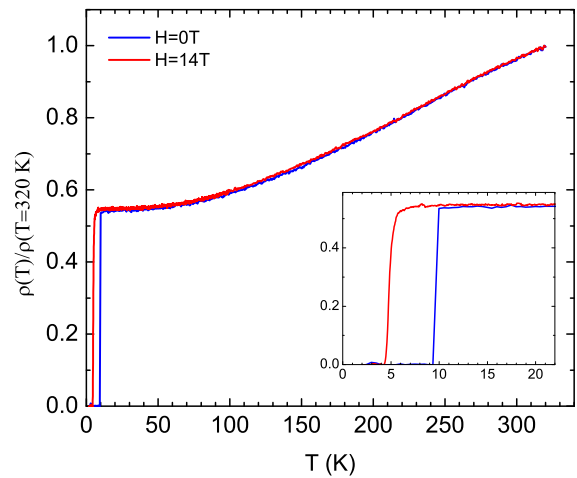}
\caption {Resistivity vs. temperature of the SrFe$_{2-x}$Ni$_{x}$As$_{2}$, x=0.155 normalized at its value at 320 K, in zero field and in 14 T//$c$. Before the measurement, the sample was annealed in low argon pressure for 24 hours at 700$^\circ$ C.}\label{fig4}\end{center}\end{figure}

Figure 6 shows resistivity vs. temperature of the SrFe$_{2-x}$Ni$_{x}$As$_{2}$, with x=0.155 normalized at its value at 320 K value, in zero and in 14 T//$c$ (it is known that for H//$c$ the critical field is lower than for H//(a b) configuration \cite{Ni-doped Sr122}). Before measurement, the sample was annealed in low argon gas pressure for 24 hours at 700$^\circ$ C \cite{Rotundu1,Paglione}. Although T$_{c}$ was suppressed from 10 to 5 K, there is no signature of any induced AFM in the 14 T//$c$ data. It is important to mention that Ni-doped samples with T$_{c}$ of 5 K are well into the coexistence of Sc and AFM region on the phase diagram, therefore exhibiting robust AFM. Our high field resistivity data are in agreement therefore with neutron measurements in a 13.5 T//$c$ field on optimally electron-doped BaFe$_{1.9}$Ni$_{0.1}$As$_{2}$ that showed that the field did not induce static AFM order \cite{Zhao}.
Therefore, part of the results of the hole- and electron-doped 122s seems to be consistent with a competing static AFM order and SC, similar to that for cuprate HTSc. The high field results reported here on the hole-doped Ba$_{0.83}$K$_{0.17}$Fe$_{2}$As$_{2}$ are similar to the case of cuprates for which the AFM order is strengthened with application of a magnetic field \cite{Lake,Khaykovich}. Despite the resemblance of the shape of the phase diagrams \cite{Kivelson,Uemura} for both iron pnictide and cuprate HTSc, the superconductivity in these materials appears to be of a different nature. We believe that these results will stimulate further exploration.

\section{Summary}

In summary, in the present article we report complementary elastic neutron scattering in zero and 14.5 T and magnetization measurements in zero and 14 T on under-doped SC x=0.17, and zero field on non-SC x=0.09. While for the non-SC x=0.09 sample the AFM transition is sharp consistent with a weakly 1$^{st}$ order transition, for higher doping x=0.17 the transition presents a broad distribution on T$_{N}$ due to a slight variation of the K dopant. For sample x=0.09 the temperature dependence of the magnetic OP is more gradual than for the case of parent BaFe$_{2}$As$_{2}$, indicative of proximity to a tricritical point. This tricritical point seems to be an universal feature among all superconducting 122s.
The slight variation on the K dopant in the x=0.17 SC sample contributes to the fractional SC volume. Although the SC under-doped x=0.17 sample has a SC volume fraction of $\approx$40$\%$, we were able to observe a downturn in the AFM order parameter below T$_{c}$, a clear sign of competition between AFM and SC, and similar to the one observed in the electron-doped 122s. As for the case of electron-doped 122s, a 14.5 T magnetic field enhances the AFM below T$_{c}$ with $\approx$10$\%$. This points, for the case at least of the 122s, toward a s$^{\pm}$ SC pairing symmetry \cite{Fernandes} in the hole-doped material, similar to that in the electron-doped 122s.
Finally we mention that while writing the current paper, we became aware of related work by E. Wiesenmayer $\emph{et al.}$ \cite{Wiesenmayer}. Their combined X-ray and muon spin rotation on powder samples of the potassium under-doped materials show microscopical coexistence of AFM and SC.

\begin{acknowledgments}

We thank Shigeru Kasahara for providing the resistivity data for the phosphorus-doped BaFe$_{2}$As$_{2}$ and E. D. Bourret for advice on the crystal growth. This work was supported by the Director, Office of Science, Office of Basic Energy Sciences, U.S. Department of Energy, under Contract No. DE-AC02-05CH11231 and Office of Basic Energy Sciences US DOE DE-AC03-76SF008. ORNL neutron scattering user facilities are sponsored by the Scientific User Facilities Division, Office of Basic Energy Sciences, U.S. Department of Energy.

\end{acknowledgments}


\begin{thebibliography}{34}
\bibitem{K41}
Since sample x=0.41 did not show static antiferromagnetism at the lowest temperature, we will present in the following only the sample with doping
x=0.09 and 0.17.
\end{thebibliography}
\end{document}